Deformation mechanism of WC single crystals under nanoindentation: Effects of surface defects and orientation on pop-in and hysteresis

H. Zhang [a]*, F. De Luca [a], H. Wang [b], K. Mingard [a] and M. Gee [a]

[a] National Physical Laboratory, Hampton Road, Teddington, TW11 0LW, UK

[b] Key Laboratory of Advanced Functional Materials, Beijing University of Technology, Beijing 100124, China

* Corresponding author: Hannah.Zhang@npl.co.uk

**Abstract**

Nanoindentation was carried out on pure tungsten carbide (WC) on the basal (0001) and prismatic (1010) planes, using Berkovich and spherical indenters, in both single load and multi-load testing. The work focuses on correlating the load-displacement curves, including elastic to plastic deformation, size effect and hysteresis with the deformation behaviour of WC. With different specimen preparation processes, the elastic to plastic deformation started at different threshold loads: This observation was found to be due to the variation in surface dislocation density. Staircase deformation was observed thought to be caused by dislocation motion and the formation of slip bands; sudden displacement discontinuities in the load-displacement response - associated with dislocation loop nucleation - occurred at, or near the theoretical shear strength. Furthermore, discontinuities in load-displacement curves were also used to confirm that hysteresis loops were a result of plastic deformation, as they when the loading was purely elastic.

## 1 Introduction

Tungsten carbide (WC) has a simple hexagonal crystal structure (with a=0.2906 nm and c=0.2837 nm, type P-m2); and is known for high melting point, chemical inertness, oxidation resistance, excellent electric conductivity, and mechanical properties [1, 2]. Pure WC which could be sintered under high pressure and temperatures [3], is a promising candidate for shielding applications such as in fusion reactors [4]; and micro-milling/cutting [5, 6], for which the specific mechanical properties of individual WC crystals are required. Traditionally, tungsten carbide-based composite (WC as matrix and ductile metal as binder) is commonly used as drilling tools, wear-resistant parts, cutters, etc. In the composite, plastic deformation occurs in both the WC and metal binder phases. Previous research has studied the local properties of individual WC grains[7-15], such as hardness, elastic modulus, and fracture strength measurements, but also has used post-mortem imaging to investigate plastic deformation mechanisms. Significant mechanical anisotropy has been reported for WC grains at the micro- and nano-scales, with $\{10\bar{1}0\}$ the primary slip plane at all orientations.

WC has a good combination of mechanical properties; high elastic modulus (close to diamond in basal orientation) [16, 17], and high hardness with plastic flow at room temperature [11, 13, 14]. Roebuck et al [18] reported varying degree of plasticity in WC grains (observations made on extensively strained WC-Co composite, i.e. 11 %); while some grains exhibited significant plastic deformation (large amount of dislocations), others remained elastically deformed or cracked with limited plasticity. In order to shed light on this plastic deformation behaviour and failure mechanisms of WC, micropillar tests [9, 19] have been carried out to measure stress-strain/displacement curves, yield and fracture strength. Csanadi et al [9] conducted micropillar compression tests of WC single crystals and found a strong influence of crystal orientation on rupture strength (12.5 ± 1.7 and 7.2 ± 0.8 GPa for basal and prismatic oriented planes, respectively) but not on yield strength (6.6 ± 0.8 and 6.3 ± 1.0 GPa for basal and prismatic oriented planes, respectively). The yield strength was much higher than the stress of deformed WC grains experienced in composite. Hence, it is of interest to study the transition from elastic to plastic deformation of WC crystals, as well as the associated dislocation activities.

Hysteresis deformation in WC-Co composite was first reported by Almond et al [18]; an attempt was made to investigate the cause of the hysteresis effect, *via* repetitive loading-unloading cycles of WC composites under compression. The cobalt binder phase was initially thought to be responsible for the hysteresis behaviour, due to its lower Young's modulus. The anelastic behaviour allows for damping/dissipation of mechanical energy generated in fatigue or during fracture, which is particularly relevant in applications such as in precision machine tools, quiet, vibration-free machinery and transportation equipment. Owing to recent developments in characterisation techniques at small scales, and their ability to test individual phases in composites materials, the phenomenon of hysteresis has been explored further. For instance, the hysteresis deformation on MAX phase ceramics and other materials has successfully been investigated by nanoindentation[20-22].

In this paper individual WC crystals were indented using multi-loading modes and investigated their hysteresis deformation behaviour. The importance of the initial dislocation density and distribution of mobile dislocation segments at yield were investigated.

**2 Experiment and materials**

2.1 Materials and characterisation

To investigate the effect of dislocation density on nanoindentation, the samples were prepared with different surface finishes. First, pure tungsten carbide (few millimetres in size) with different orientations were progressively mechanically polished with SiC paper, followed by diamond suspensions, with a final finish using silica polishing suspension (OPS) for 10 min. After indentation were carried out on OPS finished samples, the WC crystals were further polished using an ion polisher at an acceleration voltage of 6 V, 3 V and 1.5 V for 90, 90 and 60 min, respectively. WC crystal orientations were identified using a scanning electron microscope (SEM, Zeiss Supra 55, Carl Zeiss, Jena, Germany) equipped with an electron beam scanning diffraction (EBSD, Oxford instrument). The morphology of indents was investigated by SEM (secondary electron imaging), and atomic force microscopy using a tapping mode (AFM, Parker, XE 100, Suwon, South Korea); AFM scans were analysed with Gwyddion software.

2.2 Nanoindentation

2.2.1 Indenter

Berkovich and coni-spherical (6 µm radius) diamond indenters were used in this work. A 2-µm pitch silicon saw-toothed artefact calibrated AFM was used to determine the tip shape of the Berkovich indenter, over an area of $2 \times 2$ µm, using $512 \times 512$-pixel scan resolution; high resolution scans were carried out to accurately determine the apex of the indenter. Figure 1(a) shows the Berkovich indenter apex topography acquired by AFM. This showed an indenter radius of about $500 \pm 20$ nm along the first 40 nm of the indenter tip. The radius of the spherical indenter was also measured by acquiring SEM micrographs. The area function of the Berkovich indenter at higher depths was indirectly determined by making indents in reference materials (fused silicon and tungsten).

2.2.2 Indentation procedure

Indents were performed with two nanoindenter systems, one equipped with a Berkovich indenter (UHNT Anton Parr, Austria), the other with a coni-spherical indenter and (NanoTest Xtreme, Micromaterials, UK), in a relative humidity and temperature-controlled room (ambient conditions). The effect of indentation size on the mechanical properties (elastic modulus and hardness) and hysteresis behaviour of OPS-polished WC was investigated with a diamond Berkovich indenter. Ten individual indents with 2 or 3 constant load cycles were performed at each maximum load, namely, 1, 2, 3, 5, 10, 20 and 100 mN. A fixed loading and unloading time of 30 s was used along with a hold segment at maximum load of 60 seconds to allow for creep. Displacement data were corrected for thermal drift through a 60 s holding period at the end of each indent. The degree of hysteresis was characterised by the ratio of the area within the hysteresis loop) over the overall area of the load-displacement curve (total plastic work) (Figure 1(b)); the size of the loop is a measure of the reversible energy dissipated during each cycle, as described in previous work [20, 23, 24].

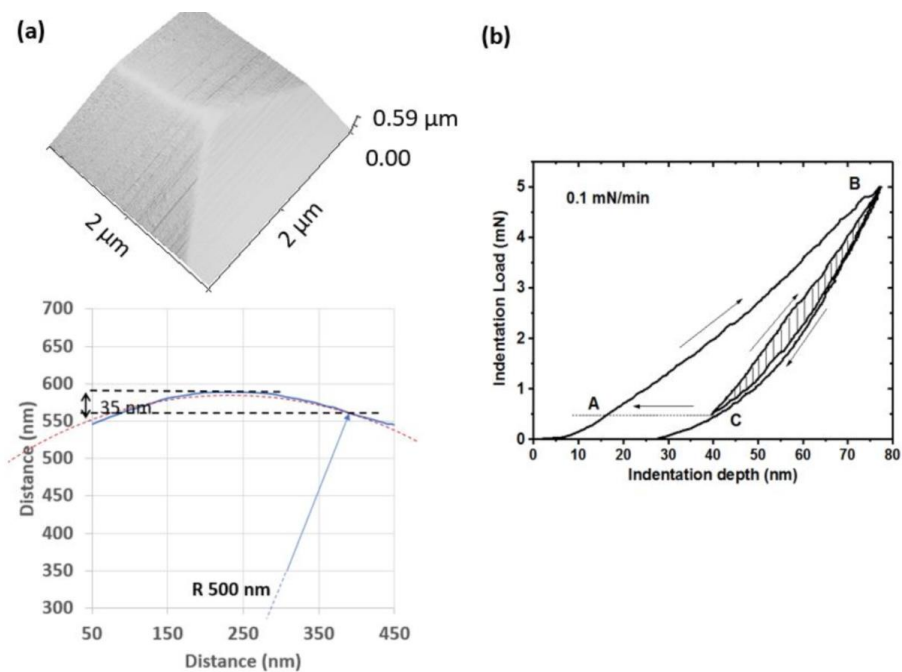

**Figure 1. Determination of indenter radius:** (a) 3D AFM micrograph of a Berkovich indenter and a representative cross-section profile of the apex(b) Load-displacement nanoindentation curve acquired on a prismatic WC crystal, exhibiting a hysteresis loop after re-loading.

To investigate the effect of surface dislocations on the mechanical behaviour of WC crystals, indents with both Berkovich indenter and coni-spherical indenter were made on an Ar ion-polished WC crystal. A grid of indents with Berkovich indenter (3x3 indents - 10 µm spacing), using a multi-loading mode (x 15 cycles) with an increasing load up to 20 mN, were carried out in an attempt to study the indentation size effect and obtain indentation strain vs indentation stress curve. Additionally, the effect of loading and unloading rate on load-displacement curves - using 0.1, 1, 5, 10 and 100 mN/min rates and a maximum load of 5 mN - was investigated by performing 5 indents at each loading rate on two crystal orientations. With the coni-spherical indenter, a multi-loading indentation with increasing loads up to 150 mN (4 indents, x 16 cycles) and 400 mN (6 indents, x 30 cycles) on basal and prismatic WC crystals, respectively, was carried out at a constant loading and unloading rate of 5 mN/min. All indents were analysed using the Oliver & Pharr method [25] to determine hardness and modulus. To give an understanding of deformation mechanisms through post-mortem AFM imaging 6 µm spherical indents were conducted on WC crystals generating radially symmetrical plastic deformation around the indent.

## 3 Results

3.1 Effect of surface finish on load-displacement curves

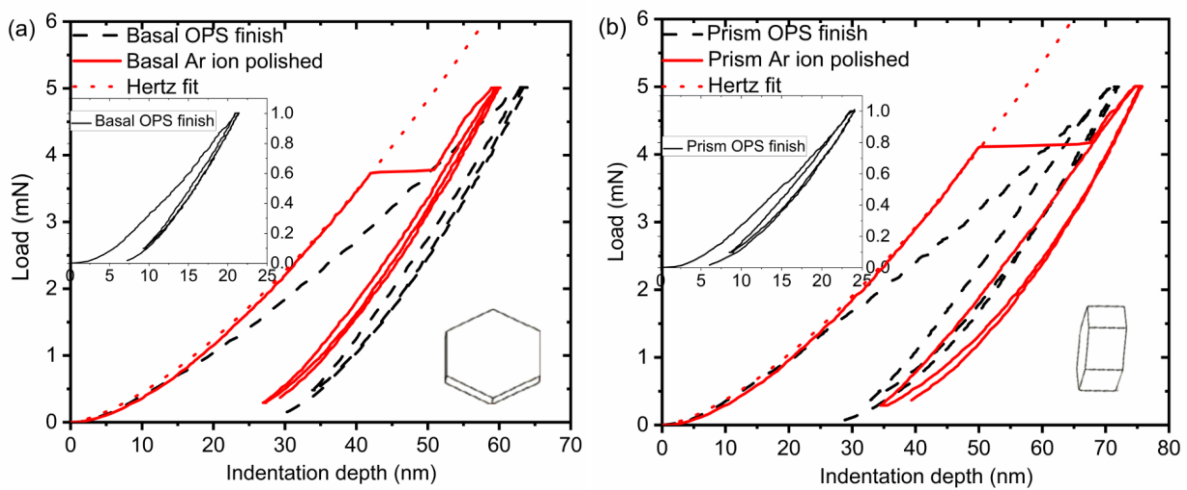

**Figure 2. Load-displacement curves of Berkovich indents**: (a) WC basal plane and (b) WC prismatic plane with different surface finishing.

Figure 2 shows load-displacement curves of Berkovich indents performed on OPS and Ar ion-polished WC single crystal surfaces with basal and prismatic orientation. The load-displacement curves of indented OPS-polished WC surfaces show a smooth transition from elastic to plastic deformation, accompanied with small pop-ins which are more pronounced for the prismatic orientation; a divergence from the Hertzian solution, which is representative of the onset in plastic deformation, occurred at a load < 1 mN for both WC orientations (insertion in Figure 2). In this figure, the first pop-in happens at a load of 0.66 ± 0.12 mN and 0.62 ± 0.14 mN, for the basal and prismatic WC orientation, respectively, regardless of the loading rate. In contrast an abrupt elastic to plastic transition was observed for Ar ion-polished WC surfaces through a single and large pop-in which was found to be independent of loading rate. When the results from multiple indents were examined, some variation in the load at which the pop-in occurred. This can be attributed to local changes in the surface properties of WC (dislocation density, residual stress, etc.). Pop-ins were also found to occur on unloading or during the loading segment of the second cycle (in case of a purely elastic first cycle).

By unloading and re-loading to the same maximum load multiple times, WC crystals exhibited hysteresis behaviour over these repeated unloading/reloading cycles. Indentation hysteresis behaviour previously been observed for other materials [21] and has been reported as an intrinsic material property rather than an instrumentation-related artefact.

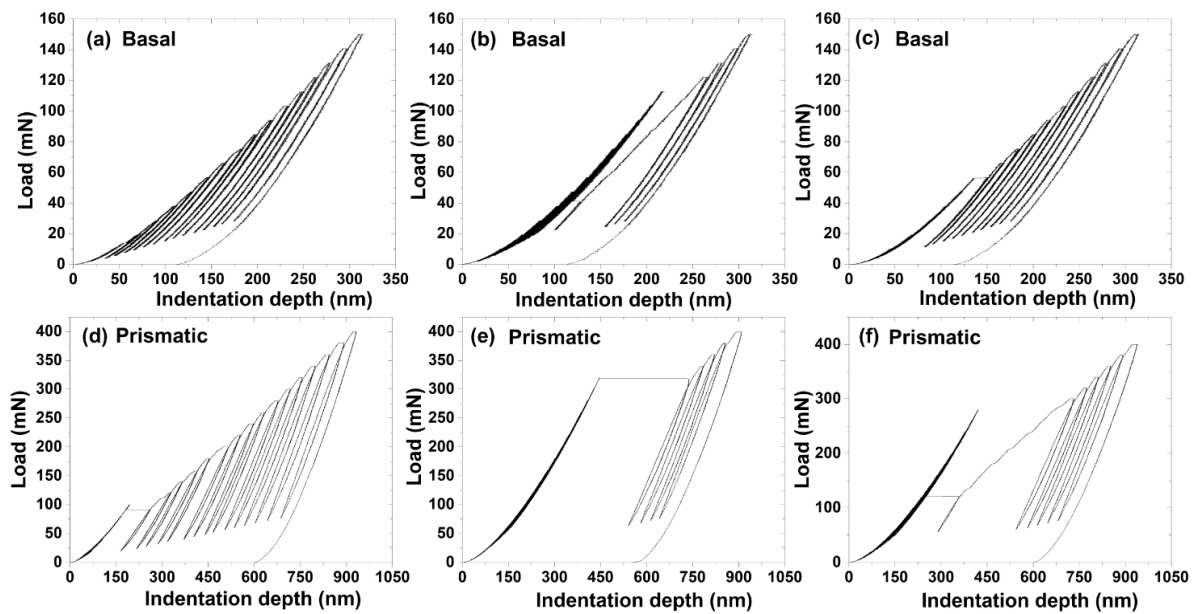

**Figure 3. Load-displacement curve for different loading profiles showing elastic-plastic transition and hysteresis behaviour in WC crystals**: Spherical indentation (R=6 µm) in basal (a-c) and prismatic (e-f) WC crystal after ion polishing.

Pop-in events and hysteresis loops were also observed in the spherical indentation experiments (R=6 µm). Figure 3 shows incremental multi-loading load-displacement curves at different locations carried out on the Ar ion-polished WC crystals. Gradual elastic to plastic transitions (Figure 3(a)) as well as more abrupt plastic formation *via* the occurrence of pop-ins either upon unloading from a nearly pure elastic deformation (Figure 3(b)) or at maximum load during holding (Figure 3(c)) were observed for basal WC crystals. By contrast sudden displacement bursts were observed for prismatic WC crystal, at random loads, either upon holding (Figure 3(e)) or unloading (Figure 3(d) and (f)). Purely elastic loading was observed prior to the formation of the sudden burst and, therefore, as a result of stress accumulation under the indenter, the magnitude of the burst was found to scale with the load at which it occurs. Whilst these large pop-ins forming upon unloading were observed for both spherical and Berkovich indentation, they have not previously been reported. Hysteresis loops were not found during elastic deformation, which suggests that it is related to the formation of dislocations and more especially their spatial distribution and mobility upon change in the stress-state of the material.

3.2 Mechanical properties

Due to the mechanical anisotropy of WC crystal, all elastic constants of WC were used as previous reported in literature [16, 17]: based on the literature values of C11, C12, C13, C33, C44, theoretical young's moduli for the basal and prismatic orientation of WC were calculated to be about 925 GPa and 620 GPa, respectively. Poisson's ratios were calculated to be $v_{prism} = 0.33$ and $v_{basal} = 0.1$ respectively. Based on the standard Oliver-Pharr evaluation method, Young's modulus and contact pressure values are calculated related to crystal orientation (as suggested in Ref [29]), and values extracted from different contact depths are presented in Figure 4.

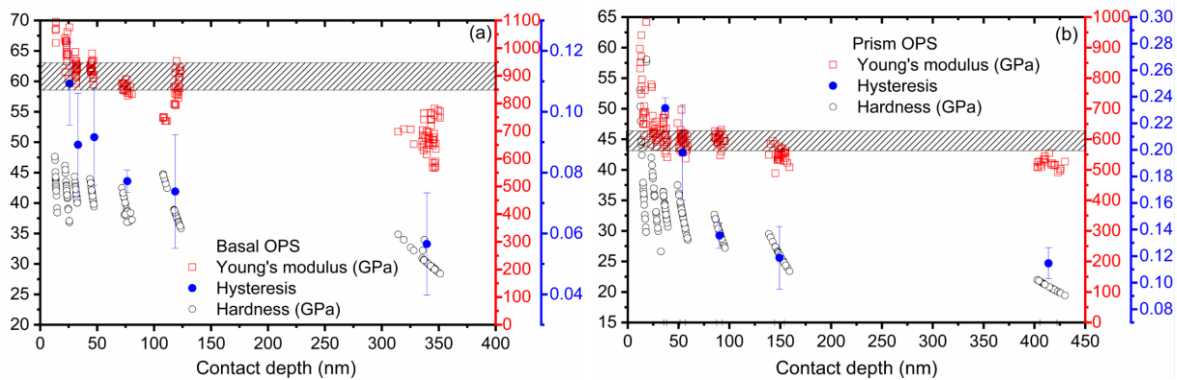

**Figure 4. Mechanical properties and hysteresis loop:** hardness, elastic modulus and size of the hysteresis loop as a function of the contact depth for OPS-polished basal (a) and prismatic (b) orientations indented with a Berkovich indenter - the dashed areas represent the theoretical elastic response with a 5 % tolerance.

The Young's modulus of WC crystals were measured about 900 ± 45 GPa and 600 ± 30 GPa for the basal and prismatic orientation (as shown in Figure 4), respectively, which is consistent with previous theoretical calculation [16, 17] and experimental results [9, 11]. Whilst a reduction in the mechanical properties of WC was observed at high contact depths (> 300 nm), indicative of cracking [26], some scatter in the data acquired at contact depths lower than 25 nm was attributed to surface roughness. The magnitude of hysteresis formed at the onset of plasticity was found to rapidly decrease as more plasticity was produced under the indenter for both WC orientations; though a higher degree of hysteresis was measured for the prismatic orientation compared to the basal orientation across the entire range of contact depths. Although the ratio of hysteresis deformation to plastic deformation energy reduced with contact depth, the absolute dissipated energy of hysteresis deformation increased.

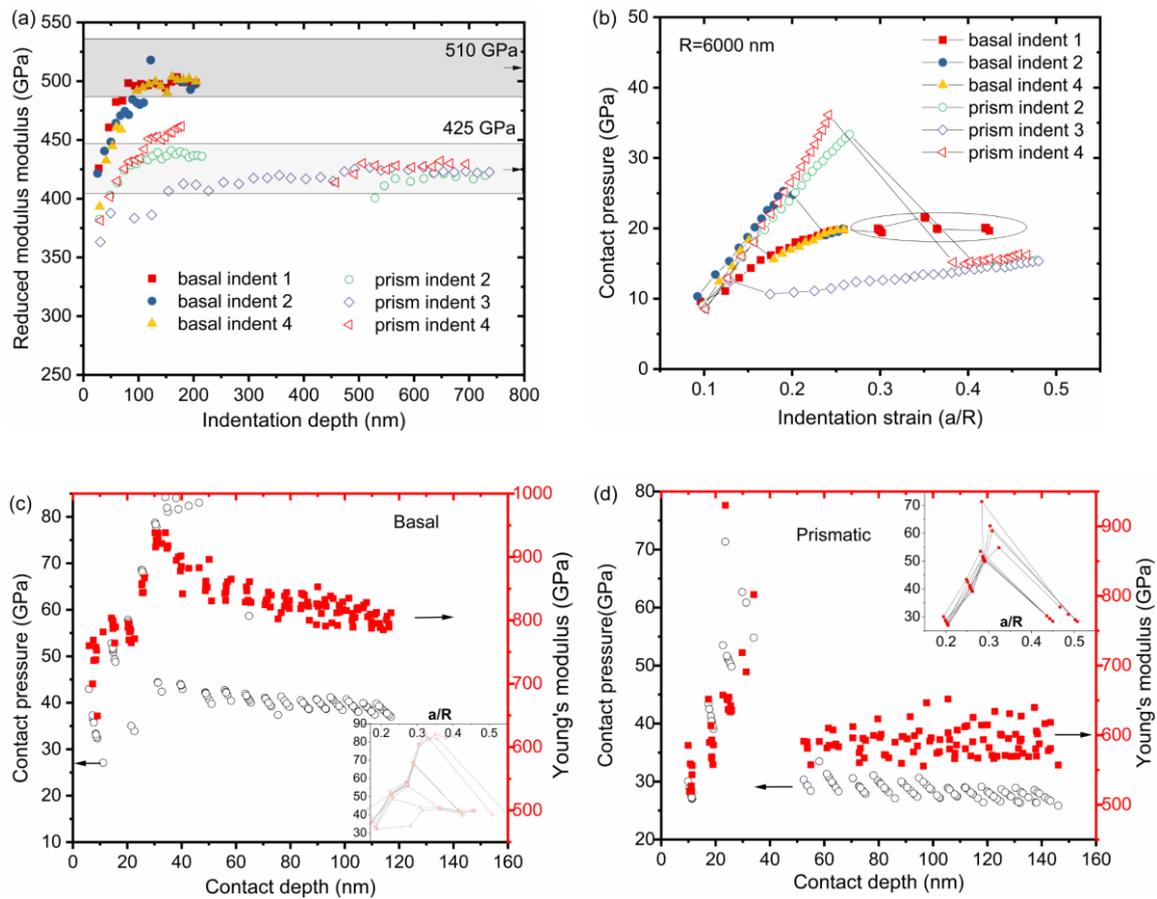

**Figure 5. Contact pressure and elastic modulus of Ar ion-polished WC crystals:** Analytical results of spherical indents from load-displacement curves shown in Figure 3 (a and b), the data marked by the ellipse in 5(b) is obtained from single load tests; Berkovich indents in basal and prismatic WC crystal (c and d, respectively)- indentation strain represented by the *a/R* ratio (according to Ref. [26]).

Spherical indentation allows for the measurement of indentation stress and indentation strain values and thus the evaluation of the elastic to plastic transition through the derivation of stress-strain curves[26]. For WC the difficulty lies in characterising initial yield behaviour and defining a reproducible yield point because of discontinuous yield phenomenon such as 'pop-in' and sudden divergence from an Hertzian contact. For this reason, measurements were made on Ar ion-polished WC crystals, where the large pop-in can be triggered by both Berkovich and spherical indents, using the Oliver & Pharr method to obtain indentation stress-strain curves. Based on Figure 3, the reduced modulus as a function of indentation depth, and indentation stress-strain curves were plotted in Figure 5(a, b). With a similar cycling load, the contact was spherical with Berkovich indents at contact depth lower than 50 nm, allowing for the spherical contact area to be converted into indentation strain, as presented in Figure 5 (c, d).

In contrast to the OPS polished surface (Figure 4), where elastic modulus is nearly constant despite of the scattering at the lower contact depth and decrease at highest contact depth, the elastic modulus of Ar ion polished sample measured from both indenters increased until it reached a plateau at contact depth of ~100 nm and 40 nm for spherical and Berkovich indenter, respectively (Figure 5). Similar phenomena were observed in single crystal MgO, which raised concerns about interpretation of unloading curves of materials with anisotropic elastic properties[27]. This was mainly due to the application of the Oliver-Pharr method which was originally developed to analyse isotropic elastic-plastic unloading curves; for the analysis of anisotropic pure elastic deformation region of WC crystals it is recommended that the elastic contact analysis of Hertz is used [29].

The elastic indentation strain (estimated by the *a/R* ratio) for both WC orientations reached up to 30 %, corresponding to an uniaxial elastic strain of up to 6% based on previous work [28]. At the transition from elastic to plastic deformation of indents with large pop-ins, the contact pressure drops sharply for both orientations, and then follows similar trend as that observed for other indents without pop-ins. The yield pressure measured from Berkovich indents was found to be much higher than that measured from spherical indents, which is proportional to the inverse square root of indenter radius in agreement with the size effect displayed in ceramics such as alumina [29].

The hardness increases with the decreasing indentation depth, which is consistent with the frequent experimental observation that 'smaller is harder' for a Berkovich indenter [30]. For spherical indenters the indentation hardness displays the opposite depth dependence with an increase of hardness with increasing indentation depth. Similar phenomena have been observed experimentally for different materials, and a simple model was developed to explain this opposite depth dependence for sharp and spherical indenters [30]. It showed that for a sharp indenter, the average plastic strain underneath the indenter is essentially independent of the indentation depth, but the plastic strain gradient is proportional to inverse of indentation depth; for a spherical indenter the plastic strain gradient becomes independent of indentation depth but the average plastic strain increases together with indentation contact radius. This leads to the opposite depth dependence of indentation hardness for Berkovich and spherical indenters when the Taylor dislocation model is considered.

3.3 Indentation tomography

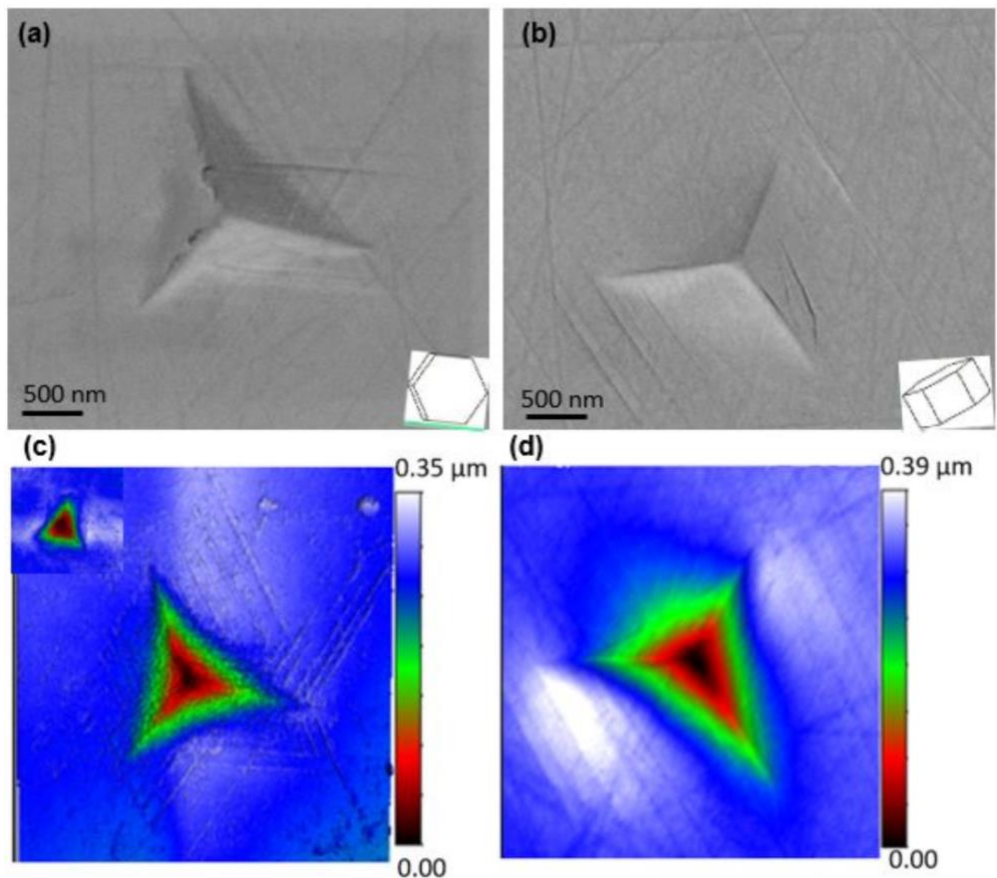

**Figure 6. Berkovich indent topography:** SEM and AFM micrographs of indents (100 mN) made in basal (a, c) and 5 mN AFM scan as inserted in (c); and prismatic (b, d) WC crystals, respectively.

The morphology of the indents was characterised by SEM and AFM, as shown in Figure 6. On the basal plane, multiple slip traces forming two reversed triangles representative of the hexagonal unit cell of WC was observed by SEM; slip traces were found parallel to the all edges of the WC hexagonal unit cell. AFM line profile shows no noticeable sink-in effect along the indent edges made on the basal plane. Both SEM and AFM images show the symmetric shape of indents (Figure 6(a, c)). On the prismatic plane, parallel intersecting slip traces were activated, along with pile-up formation as observed by AFM. The slip traces are parallel to the sides of the WC hexagonal unit cell. While the indent shape from the SEM micrograph appears to be symmetric (Figure 6(b)), the AFM scan (Figure 6(d)) shows some asymmetry in the prismatic oriented indent which is consistent with previous work [11], showing pile-ups caused by slip planes. Furthermore, micro-cracks were observed on the top surface SEM micrographs of both WC orientations after indentation at 100 mN, while no slip trace or surface micro-crack was found after indentation at a load of 5 mN.

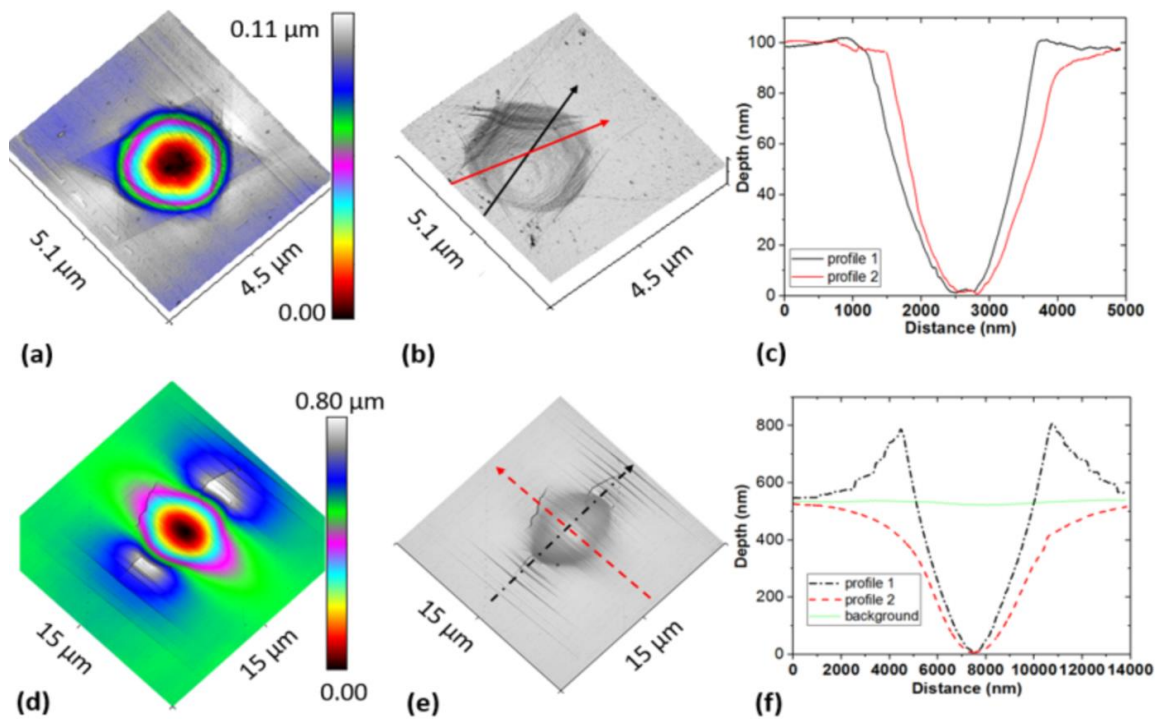

**Figure 7. Spherical indent topography:** AFM micrographs and scans of indentations made at a load of 150 mN in basal (a, b, c) and 400 mN in prismatic (d, e, f) planes - Ar ion-polished WC surfaces.

Slip line traces were also observed when indentations from spherical indentation experiments were examined. In these experiments a radially symmetrical stress field was generated as shown in Figure 7. Figure 7(a, b) shows a clear six-fold symmetry in the formation of the slip traces on the basal plane, forming a star-shape. In addition, the AFM line profiles (Figure 7(c)) of the indent do not show large sink-in or pile-up at the indentation load used. The observation is consistent with previous research showing the dislocations glide downwards along the $\{10\bar{1}0\}$ prismatic slip planes and accommodate the volume change created by the indent. Figure 7(d, e) shows a two-fold symmetry in the slip traces produced on the prismatic plane, along with the formation of pileups where upwards gliding of dislocations occurred. Along the direction of the slip plane, and across the indent, sink-in was observed on the edge of the indent. Both pile-up and sink-in were observed on the prismatic plane; this is consistent with previous simulation work on slip magnitude for materials exhibit high hardness with one soft slip system[31]. The research shows that tensile stress develops when slip is restricted, corresponding to sink-in region.

In summary, these nanoindentation experiments shed further light on the mechanical behaviour of WC:

(1) The influence of pre-existing surface defects (mainly dislocations introduced by mechanical polishing) has been demonstrated in the load-displacement curve which promotes early plastic deformation; also, there was an abrupt increase in displacement when indenting Ar ion-polished surfaces.

(2) Hysteresis deformation behaviour was observed right after the onset of plasticity on both basal and prismatic WC crystals, which was found to be more significant on the prismatic plane; the degree of hysteresis decreased with an increase indentation load.

(3) AFM scans of the indents showed clear slip patterns and both sink-in and pile-up were present in the unique depth profile of an indent made in the prismatic plane.

## 4 Discussion

For WC crystals with different surface finishes factors such as pop-ins or 'staircase yielding' associated with local variation in the number of dislocations and elastic to plastic transition are caused by dislocation nucleation or motion. For indenters with different radii at small indentation depth, increasing radius enlarges surface contact area for a given indentation depth and shows wider range of variation of load where large pop-in occurs. When a significant pop-in happens, the formation of microcracks or slip bands usually occurs as a mechanism to release the accumulated stresses. To understand the deformation mechanism or crack formation in WC crystals under the indenter it is necessary to know the material stress state during loading and compare it with its theoretical properties.

## 4.1 Deformation mechanism

### 4.1.1 Elastic regime

The initial elastic contact during indentation can be interpreted using Hertz contact theory where the load $F$ required for displacing a sphere with radius $R$ in an elastic material is given by the following equation [32]:

$$F = \frac{4}{3} E_r \sqrt{R} h^{1.5} \qquad (1)$$

where, $E_r$ is the reduced modulus of WC basal and prismatic plane (510 and 425 GPa, respectively), and $h$ is the indentation depth. The initial Hertz fit based on Eq.(1) is shown in Figure 2, with a Berkovich indenter radius of 490 nm; $h^{1.48}$ was used instead of $h^{1.5}$, slightly due to a non-perfect spherical shape of the indenter apex (as shown in Figure 1). According to previous work, the Hertz contact theory was demonstrated to apply to different materials with a/R ratio up to 0.3 or even higher [29, 33], which is close to the indentation strain where large pop-in were observed to occur in WC crystals. The latter supports our previous observation that WC crystals elastically deform prior to the first pop-in. High indentation strains, up to 30 %, for both basal and prismatic orientations correspond to uniaxial strains of about 6-9 % according to different constrain factors of conversion[28]. The yield strain of Ar ion-polished WC crystals was found to be much higher than those for OPS-polished WC crystals but varied slightly at different locations. The latter shows that surface defects such as dislocations and stacking faults influence the yield strain. Comparing the Hertz fit with the unloading curve (as shown in Figure 2, which are not overlap), indicates that further plasticity occurs during the unloading of the indentation, which is due to the residual stress field of the indentation.

### 4.1.2 Dislocation motion

Peierls stress, $\tau_p$, is the minimum external stress required to move at dislocation within a plane of atoms. For all practical purposes, it is equivalent to the critical resolved shear stress (except for thermally activated processes), which is the shear component of an applied tensile or compressive stress resolved along a slip plane other than perpendicular or paralleled to the stress axis:

$$\tau_p = \frac{2G}{1-v} \exp\left[-\frac{2\pi a}{(1-v)b}\right] \qquad (2)$$

where, *a* is the interplanar spacing of the slip plane and *b* is the interatomic spacing along the slip direction, $v$ is the Poisson's ratio. According to recent work [12, 13], the shear stress needed for dislocations to move along the most common slip systems in WC are given in Table 1.

Table 1. Peierls stresses for the different slip systems of WC.

| slip plane | slip direction | a (nm) | b (nm) | stress (GPa) |
|---|---|---|---|---|
| {11$\bar{2}$0} | <0001> | 0.1258 | 0.2837 | 21.01 |
| {10$\bar{1}$0} | <$\bar{1}$2$\bar{1}$3> | 0.2517 | 0.4061 | 5.18 |
| | <$\bar{1}$2$\bar{1}$0> | | 0.2906 | 0.73 |
| | <0001> | | 0.2837 | 0.62 |
| {0$\bar{1}$00} | <11$\bar{2}$3> | 0.2030 | 0.4061 | 13.43 |
| | <0$\bar{1}$10> | | 0.2906 | 2.77 |

While all WC slip systems described in Table 1 have already been observed by TEM imaging, the {10$\bar{1}$0}<$\bar{1}$2$\bar{1}$3> slip system is considered to be the main system of slip at room temperature by many researchers. On the same slip plane {10$\bar{1}$0}, undissociated dislocations along <$\bar{1}$2$\bar{1}$0> and <0001> were observed in areas far from an indent where sample was only deformed a small amount [34, 35]. According to Table 1, these two slip systems require much lower shear stress to glide. For example, the shear stress required for a dislocation to move along {10$\bar{1}$0}<$\bar{1}$2$\bar{1}$0> systems are as low as ~0.7 GPa; this explains the early formation of plastic deformation in OPS-polished WC crystals even for indentation load less than 1 mN (Figure 2).

Plastic deformation is believed to occur by the glide of pre-existing dislocations introduced by mechanical polishing in the vicinity of the plastic zone. After ion polishing, the deformation zone was removed in areas where scratches were introduced during mechanical polishing, and the first pop-in initiated plastic deformation through homogeneous nucleation of critical defects such as dislocation slip bands in an undeformed area.

4.1.3 Formation of slip bands

Experimental investigation of theoretical strength became possible with the emergence of small scale testing methods such as nanoindentation and micro-mechanical testing [36] where defects are absent or their concentration is negligible. These techniques allow for the testing of interatomic bonds, which either break (formation of cracks) or re-form (development of a shear). In WC with small dislocation densities (such as Ar ion-polished surfaces) dislocation nucleation occurred at small indentation depths where pop-in was found. At nanometre scales the theoretical strength is seen. Assuming a Hertzian contact, the maximum shear stress (principal shear stress) under a spherical indenter (R>>a) at a load P, where the *P-h* discontinuity occurs, approximates C* $P_0$ (maximum pressure, C=0.357 and 0.304, for ν = 0.1 and ν =0.33) at a depth of 0.5a (radius of contact area)[37]:

$$\tau_{max} = C \left(\frac{6PE^{*2}}{\pi^3 R^2}\right)^{1/3} \qquad (3)$$

where *R* is the radius of the tip of the indenter, *E\** is reduced modulus.

With Berkovich indenter, the ranges of maximum shear stresses in single WC crystals were calculated, based on a statistical distribution of loads at which pop-in events occur. The average load at which the large pop-in event occurs in Ar ion-polished WC basal and prismatic crystals was measured about 3.83±1.14 and 3.87 ± 1.24 mN, respectively, (from 25 and 20 individual indents, respectively), which corresponds to a shear stress of 33.1 ± 2.80 and 25.05 ± 2.74 GPa (with Eq. (3)).

According to previous work [38], the shear stress necessary for homogeneous dislocation nucleation can be calculated from the elastic self-energy stored in a dislocation loop. The maximum of the above relation is the critical stress $\tau_c$ at the critical loop radius $r_c$ for dislocation nucleation, $\tau_c \approx \frac{Gb}{2\pi r_c} \approx \frac{G}{10}$, where G is the shear modulus, b the magnitude of the Burgers vector, and $r_c$ the inner cut off radius. For the different orientations, (different shear modulus, poison's ratio, and young's modulus) the critical shear stress was calculated to be about 32.8 GPa and 23.3 GPa for basal and prismatic orientation, respectively; a shear modulus of 328 GPa and 233 GPa for basal and prismatic plane was used, respectively [16].

In conclusion in experiments with a Berkovich indenter on Ar-ion polished WC crystals, plastic deformation was triggered (pop-in) at a load where the critical shear stress was close to the theoretical strength of WC, through homogeneous nucleation of the first dislocation loop and subsequent avalanche multiplication. This conclusion is supported by observation of slip bands on AFM micrographs (Figure 7) and dislocation density calculation of 1 dislocation per ~0.25 (prism) and ~0.5 μm^2 (basal) in Ar-ion polished WC[39]. This area is much larger than the contact area between indenter and specimen of ~ 0.12 μm^2 when the large pop-in occurred, indicating occurring of new dislocation. The dislocation density of defects introduced upon fabrication such as vacancies, dislocations, voids, etc. [13, 34, 35, 40] and material processing (Figure 6), and their distribution reduced the critical stress $\tau_c$ of materials [41]. Furthermore, the high strength and elasticity of WC single crystals is explained because small volumes of WC were tested where the number of defects (dislocations) were limited. The results offer an insight towards the optimisation of WC nanostructure design.

4.2 Size effect

A nanoindentation size effect has previously been observed in ceramics, but compare with metallic materials, with a smaller degree of increase in the yield pressure that is proportional to the inverse square root of indenter radius [29]. The mechanism behind the size effect behaviour of ceramics reported in this work was not clear as an increase in the initial yield strength could not be explained by the strain gradient plasticity theory[42]. At the time of the research [29], the size effect focused on the study of dislocation motion, rather than dislocation multiplication and initiation [33, 42].

The nanoindentation size effect in WC was investigated by through yield strength measurement. Yield strength values close to theoretical predictions were measured for Ar ion-polished WC in experiments with a Berkovich indenter. By comparison, experiments on OPS-polished WC which gave much lower yield strengths because the dislocation density was different. Due to a larger contact area at the same indentation depth and exposure to a larger number of dislocations when the indenter radius was increased the yield strength was found to decrease for both WC orientations. The hardness for Berkovich indenter decreased with increasing indentation depth and increased for spherical indenter due to different strain gradient and work hardening rate [30] for the different geometries.

The dislocation density of undeformed WC crystal is typically about 1-10 dislocations per μm, and higher density for deformed WC crystals [43]; WC grains exhibit high strength if there are either high level of dislocation (> 10 dislocations per μm) or no dislocations. Considering different length scales and dislocation density, the size effect in WC could be summarised into 3 stages. In the first stage at small scales with a low dislocation density the strength is governed by dislocation scarcity; with an increase of dislocation density the yield strength decreases. In the second stage dislocation motion dominates plastic deformation; with an increase of dislocation density the yield strength increases. Finally, the third stage is governed by is dislocation interaction and the formation of microcracks. This means that to increase the strength of WC while maintaining its ability to plastically deform the concentration of defects such as dislocation density should be reduced.

4.3 Hysteresis

The formation of a hysteresis loop in between the unloading and reloading segments of the indentation curves of WC crystals was evidence of the occurrence of mechanical phenomena such as dislocation generation and propagation during unloading (Figure 2) with a good fit with Hertzian elastic deformation for the reloading segment of the curve. Mechanical hysteresis loops were observed in different materials after the elastic to plastic transition (yield), and the mechanism leading to hysteresis varied from microcracking, residual elastic lattice strain, kink band to dislocation avalanche. As the elastic modulus to hardness ratio (E/H) reflects the ductile and brittle nature of materials, such as the ratio is usually more than 30 for metallic materials, and about 10 for ceramic materials [44]. Therefore, the hysteresis deformation mechanism in different materials are summarised in Table 2, together with material's E/H ratio.

Table 2 Hysteresis behaviour of different materials under nanoindentation.

| Mechanism | Materials | Modulus (GPa) | Hardness (GPa) | E/H ratio | Ref |
|---|---|---|---|---|---|
| **Microcracking** | Diamond (0001) | 1141 | 100 | 11.04 | [25] |
| | Fused silica | 72 | 9.2 | 7.8 | [23] |
| **Residual elastic lattice strain** | WC (0001) | 825 | 40 | 21.8 | Current work |
| | WC ($10\bar{1}0$) | 600 | 30 | 20 | |
| | MgO | 285 | 14 | 20.4 | [22] |
| | Polycrystalline $Ti_3SiC_2$ | 325 | 20 | 16.5 | [22] |
| **Incipient Kink Band** | $Ti_3SiC_2$ (0001) $Ti_3SiC_2$ ($10\bar{1}0$) | 300 | 12 | 25 | [21, 23] |
| | $Ti_2AlC$ | 268 | 6 | 44.6 | [20] |
| **Dislocation avalanche** | Tungsten | 420 | 6 | 70 | [25] |
| | Nanosized copper pillar <111> | 150 | 1.5 | 100 | [24] |

In a similar way to WC, MgO single crystals exhibit strong plastic anisotropy, and the hysteresis deformation under indentation was found using complex neutron diffraction analysis [23] to be caused by the residual elastic lattice strain. Previous work [31, 45] shows that enough independent slip systems must operate to accommodate the indentation strain through plastic flow and avoid the accumulation of elastic stresses. If this cannot occur the release of elastic stresses through unloading can trigger reverse plastic flow and the formation of reversible hysteresis behaviour[46].

This explanation for the deformation mechanism of WC single crystals is supported by the experimental observations. The micrographs (Figure 7) of the spherical indent show six-fold and two-fold symmetries for the basal and prismatic WC orientation, respectively. According to previous work [11] the anisotropic nature of the indent topography for different crystallographic orientations of WC crystals are correlated with activated slip systems. The most common slip system is $\{10\bar{1}0\}<\bar{1}2\bar{1}3>$ for WC and the associated 6 and 2 slip bands on the basal and prismatic plane respectively. Although other slip systems could be activated (Table 1), they require higher stresses. The basal orientation was found to have less hysteresis than the prismatic orientation of WC because there is a higher number of slip systems to allow for the accommodation of stress. The assumption that the hysteresis behaviour relies on the availability of slip systems is in line with the observations made with regards to size effect. The degree of hysteresis reduces with an increase in indentation depth and is likely to be due to the activation of slip systems.

## 5 Summary and conclusions

In WC crystals, the density of pre-existing dislocations determines the yield stress and the nature of yield through either dislocation propagation or generation. Mechanically polished WC crystals exhibit a high number of surface dislocations which promote the mobility of dislocations under the indenter leading to plastic deformation. After Ar ion-polishing the dislocation density on the surface of WC crystals was reduced as indicated by the occurrence of pop-in events. While performing indentation with an indenter radius of about 500 nm pop-in events were consistently observed at a specific load. In regions with reduced dislocation density the critical shear stress for the nucleating of homogeneous dislocation was found to be close to the theoretical shear stress value. By contrast pop-in events were more erratic and location-dependent when using an indenter radius of 6000 nm. The indentation size effect was also reflected by lower hardness values from indentations performed with a larger indenter radius, which could be due to different stress gradient and cracking under the indenter. Interestingly, pop-in events were observed during unloading, mimicking the fatigue failure of ceramic materials and indicative of stress accumulation. SEM and AFM micrographs show there were a limited number of activated slip systems and indicate plastic anisotropy in WC crystals promoting the development of elastic strains under the indenter with the consequence of reverse plastic flow. Experimental observations such as the higher hysteresis for the prismatic oriented plane and its decrease with increasing load shows that the lack of easy slip systems under indentation caused the hysteresis to form.

# 6 Acknowledgement

The authors would like to thank Dr Bryan Roebuck for helpful discussion and Dr Helen Jones for providing WC crystals; and the National Measurement System Programme of the UK government's Department for Business, Energy & Industrial Strategy (BEIS) for financial support.